\documentclass[%
 reprint,
 amsmath,amssymb,
 aps
]{revtex4-1}

\usepackage[dvipdfmx]{graphicx}
\usepackage{tikz}
\usetikzlibrary{arrows,snakes,backgrounds}

\usepackage{bm}
\usepackage{amsmath}\def\hsymb#1{\mbox{\strut\rlap{\smash{\Huge$#1$}}\quad}}
\usepackage{cancel}
\usepackage{braket}
\usepackage{amsthm}
\usepackage{dcolumn}
\usepackage[capitalize]{cleveref}
\usepackage[caption=false]{subfig}
\begin{document}
\preprint{APS/123-QED}
%\preprint{}

\title{Recursive Green's function approach to Feenberg perturbation theory}


\author{Kazuya Ishida}
\email{ishida-kazuya@ed.tmu.ac.jp}
%\thanks{}
\affiliation{Department of Physics, Tokyo Metropolitan University, Minami-Osawa, Hachioji, Tokyo 192-0397, Japan}


\date{May 31, 2019}

\begin{abstract}
We propose a new procedure by using the recursive Green's functions which remove all the repetition terms from the time-independent perturbation series for finite-level quantum systems. These Green's functions are introduced as a generalization of the Brillouin-Wigner perturbation theory and the calculations of their diagonal elements can naturally give the effective propagators which are equal to the ones in the Feenberg perturbation theory. 
\end{abstract}
%\keywords{}

\maketitle
\section{\label{sec:level1}INTRODUCTION}
In quantum mechanics, the time-independent perturbation theories can be classified into three types in terms of removing the repetition terms which appear in the eigenstates. The Rayleigh-Schr${\ddot{\rm o}}$dinger perturbation theory(RSPT)\cite{1,2}  has a naive perturbation series representation in which the repetition terms are exposed explicitly. In the Brillouin-Wigner perturbation theory(BWPT)\cite{3,4,5}, the repetition terms corresponding to a self-energy are removed by partially modifying the energy denominators.  

The Feenberg perturbation theory(FPT)\cite{6,7} is also known for removing all the repetition terms from the eigenstates and thus is often regarded as a generalized BWPT. It has been studied in many ways: Feshbach's successive approximation\cite{8} the continued fraction(CF) method\cite{9,10}, projection operator(PO) approach\cite{11}. However, these approaches are not systematized as a natural extension of the BWPT. In addition, the Feenberg's formulation originally has a complicated structure and the various index notations for the effective propagators or modified energy denominators have been proposed in each of the approaches. Therefore, the representations of their formulas are slightly different and confusing in actual use.

In this paper, we present the new derivation method of the Feenberg perturbation theory by using the recursive Green's functions. These Green's functions are suitable for systematically removing all the repetition terms from the eigenstates as a generalization of the derivation process of the BWPT. It is interesting that the Feenberg's perturbation series is directly derived from a simple recurrence relation which the recursive Green's functions yield and then their diagonal elements play the role of the effective propagators in the FPT. Furthermore, for finite-level quantum systems, the equation for the energy eigenvalue becomes an exact self-consistent equation formed as a finite continued fractions. This is the same result as the CF and PO approaches. In our approach, the analysis of the effective propagators can also provide some important clues for clarifying the relationships between the FPT, BWPT and RSPT, linking our approach and another derivation methods of the FPT and verifying the valid of the FPT in the strong coupling region.

This paper is organized as follows. In Sec.~\ref{sec:level2}, we discuss the removing process of the repetition terms in the BWPT in terms of Green's function method and introduce the recursive Green's functions defined on a subspace at each step to generalize the removing process. Then, we consider $N$-level quantum system in which the spectrum of an unperturbed Hamiltonian ${\hat H}_0$ is discrete and has no degenerate. The two calculation procedures for the effective propagators are shown in Sec.~\ref{sec:level3}. One is the Neumann series expansion with respects to a local self-energy and then the relationships between the FPT, BWPT and RSPT are clarified in terms of the approximation and resummation. The other is a determinant expression for themselves which rewrite the formulations given by another approaches and distinguishes between removable and non removable singularities. Sec.~\ref{sec:level4} gives several numerical calculations in the case $N=7$ to verify the validity of the formulations derived through our approach in terms of the convergence and exactness. Sec.~\ref{sec:level5} is devoted to the summary. 

\section{\label{sec:level2}GREEN'S FUNCTION METHOD}
We first consider the time-independent Schr${\ddot{\rm o}}$dinger equation written in the following form
\begin{equation}
({\hat H}_0+\lambda{\hat V})\Ket{\psi_{\tau_0}}=E_{\tau_0}\Ket{\psi_{\tau_0}},\hspace{5mm}\tau_0=1, 2, 3,\ldots N
\end{equation}
where an unperturbed Hamiltonian ${\hat H}_0$, an interaction term ${\hat V}$ and a coupling constant $\lambda$ are given. We already know the eigenvalues ${\omega}_{\tau_0}$ and eigenstates $\Ket{\tau_0}$ of ${\hat H}_0$: ${\hat H}_0\Ket{\tau_0}={\omega}_{\tau_0}\Ket{\tau_0}$, while $E_{\tau_0}$ and $\Ket{\psi_{\tau_0}}$ are the unknown eigenvalues and eigenstates of the full Hamiltonian ${\hat H}\!=\!{\hat H}_0+\lambda{\hat V}$, respectively. In addition, both sets $\{\Ket{1},\ldots\Ket{N}\}$ and $\{\Ket{\psi_{1}},\ldots\Ket{\psi_{N}}\}$ become two orthonormal bases for $N$-dimensional Hilbert space $M$. 

To find the perturbed eigenvalue $E_{\tau_0}$ and eigenstate $\Ket{\psi_{\tau_0}}$ in terms of Green's function method, we define a total Green's function ${\hat G}(z)$ on a complex plane
\begin{equation}
(z-{\hat H}){\hat G}(z)=i.
\end{equation}
This is the resolvent of ${\hat H}$ and its spectral representation is given as follows
\begin{equation}
{\hat G}(z)\equiv\frac{i}{z-{\hat H}}=\sum_{\alpha=1}^{N}\ket{\psi_\alpha}\frac{i}{z-E_\alpha}\bra{\psi_\alpha}.
\end{equation}
In our approach, we construct the eigenstate on the basis of the fact that if $\Braket{\tau_0|\psi_{\tau_0}}\neq0$, then the following limit as z approaches $E_{\tau_0}$ is the $\tau_0$-th eigenstate $\ket{\bar{\tau}_{0}}$ 
\begin{equation}
\lim_{z \to E_{\tau_0}}\frac{{\hat G}(z)\ket{\tau_0}}{\bra{\tau_0}{\hat G}(z)\ket{\tau_0}}=\ket{\bar{\tau}_{0}}.
\end{equation}
Note that this assumption is satisfied in the non-degenerate case because the perturbed eigenstate can be evaluated as $\Ket{\psi_{\tau_0}}\sim\Ket{\tau_0}$ on this condition. Furthermore $\ket{\bar{\tau}_{0}}$ is unnormalized but the $\tau_0$-th element is always 1. This formulation is commonly used in the RSPT and BWPT, and we can divide the above ket before taking the limit into an unperturbed part $\ket{\tau_0}$ and a perturbed $\ket{g'_{\tau_0}}$
\begin{equation}
\frac{{\hat G}\ket{\tau_0}}{\bra{\tau_0}{\hat G}\ket{\tau_0}}=\ket{\tau_0}+\ket{g'_{\tau_0}},
\end{equation}
where $\Braket{\tau_0|g'_{\tau_0}}=0$. To obtain the equation for $\ket{g'_{\tau_0}}$, we now utilize a projection operator ${\hat P}_{[\tau_0]}$ onto the subspace $M_{[\tau_0]}$ which is spanned by a set of all the unperturbed eigenstates except $\ket{\tau_0}$
\begin{align}
{\hat P}_{[\tau_0]}\equiv\sum_{\tau_1\neq \tau_0}^N\Ket{\tau_1}\!\Bra{\tau_1},\hspace{5mm}{\hat P}_{[\tau_0]}\ket{g'_{\tau_0}}=\ket{g'_{\tau_0}}.
\end{align}
Operating on both sides of Eq. (5) with ${\hat P}_{[\tau_0]}(z-\hat H)$ and using Eq. (2), it is given as follows
\begin{equation}
{\hat P}_{[\tau_0]}(z-{\hat H}){\hat P}_{[\tau_0]}\ket{g'_{\tau_0}}=\lambda{\hat P}_{[\tau_0]}{\hat V}\ket{\tau_0}.
\end{equation}
This is a linear equation described on $M_{[\tau_0]}$ and can be solved by introducing a recursive Green's function ${\hat G}_{[\tau_0]}$ at 1-step
\begin{equation}
{\hat P}_{[\tau_0]}(z-{\hat H}){\hat P}_{[\tau_0]}{\hat G}_{[\tau_0]}=i{\hat P}_{[\tau_0]},\hspace{3mm}{\hat P}_{[\tau_0]}{\hat G}_{[\tau_0]}{\hat P}_{[\tau_0]}={\hat G}_{[\tau_0]}.
\end{equation}
Hence, we can formally rewrite Eq. (5) in the following form
\begin{equation}
\frac{{\hat G}\ket{\tau_0}}{\bra{\tau_0}{\hat G}\ket{\tau_0}}=\ket{\tau_0}+(-i\lambda){\hat G}_{[\tau_0]}{\hat P}_{[\tau_0]}{\hat V}\ket{\tau_0}.
\end{equation}
We should note that the right-hand side of Eq.(9) does not contain any repetition terms ($\tau_0\rightarrow\tau_0$) on the Hilbert space $M$. It is obvious from that when the BW propagators are the unperturbed ones, the Neumann series expansion of ${\hat G}_{[\tau_0]}$ directly yields the BW series which is known for having no repetition terms ($\tau_0\rightarrow\tau_0$). Now, considering that a ket ${\hat G}{\ket {\tau_0}}$ is composed of all the possible virtual processes which occur in $M$, we can interpret that the division by a diagonal element $\bra{\tau_0}{\hat G}\ket{\tau_0}$ remove the repetition terms ($\tau_0\rightarrow\tau_0$) on $M$ from the ket ${\hat G}{\ket {\tau_0}}$ and thus the BW series is reproduced.

In this section, we generalize the above scheme of removing the repetition terms in the BWPT and remove all the repetition terms from the perturbation series. To see that, rearranging the right-hand side of Eq.(9) in the following form
\begin{equation}
\ket{\tau_0}+(-i\lambda)\!\!\!\sum_{\tau_{1}\neq{\tau}_{0}}^N\!\frac{{\hat G}_{[\tau_0]}\ket{\tau_1}}{\bra{\tau_1}{\hat G}_{[\tau_0]}\ket{\tau_1}}\bra{\tau_1}{\hat G}_{[\tau_0]}\ket{\tau_1}\bra{\tau_1}{\hat V}\ket{\tau_0}
\end{equation}
we can form the ket which is the same form as the left-hand side of Eq.(5). Therefore, by repeating the process from Eq.(5) to Eq.(10) for it formally, all the repetition terms may be separated as the diagonal elements of the recursive Green's functions introduced at each step.
\subsection{Recursive Green's function}
In order to perform the above iterate scheme, let us employ the notations: the principal quantum number of an initial state $\tau_0$ and the intermediate states $\tau_1, \tau_2,\ldots$ following $\tau_0$. Then, we define the recursive Green's function ${\hat G}_{[{\bm p}_k]}$ at $k$-step in the following form
\begin{equation}
{\hat P}_{[{\bm p}_k]}(z-{\hat H}){\hat P}_{[{\bm p}_k]}{\hat G}_{[{\bm p}_k]}=i{\hat P}_{[{\bm p}_k]},\hspace{3mm}k=0, 1,\ldots N-1
\end{equation}
where ${\bm p}_k$ is $k$-tuple of the variables from $\tau_0$ to $\tau_{k-1}$ and ${\hat P}_{[{\bm p}_k]}$ is a projection operator onto the subspace $M_{[{\bm p}_k]}$ which is spanned by a set of all the unperturbed eigenstates except ${\tau}_{0},{\tau}_{1},\ldots{\tau}_{k-1}$-th ones
\begin{align}
{\bm p}_k=[\tau_{k-1}\ldots, \tau_{1}, \tau_{0}],\hspace{10mm}{\hat P}_{[{\bm p}_k]}\equiv\!\!\sum_{\tau_{k}\neq{\tau}_{i}\atop i<k}^N\Ket{\tau_k}\!\Bra{\tau_k}.
\end{align}
Note that the ${\hat G}_{[{\bm p}_0]}$ at $0$-step is equal to the total Green's function $\hat G$. We also assume that ${\hat P}_{[{\bm p}_k]}{\hat G}_{[{\bm p}_k]}{\hat P}_{[{\bm p}_k]}={\hat G}_{[{\bm p}_k]}$. In the following, we discuss the properties of the recursive Green's function.

Firstly, we explain the origin of an enclosed index ${\bm p}_k$ in square brackets. This ${\bm p}_k$-dependency is generated as a by-product of the iterative scheme from $1$ to $k$-step. More specifically, at each $i$-step ($i=1, 2,\ldots k-1$), we calculate the recursive Green's function by focusing on its $\tau_i$-th column vector (See Appendix B) and thus the inserted things at $k$-step end up depending on the $k$-tuple. Then, we specify the ${\bm p}_k$-dependency of the inserted things like ${\hat G}_{[{\bm p}_k]}$, ${\hat P}_{[{\bm p}_k]}$ and $M_{[{\bm p}_k]}$. Furthermore, it can be regarded as $k$-permutations of a set $\{1, 2,\ldots N\}$ when $\tau_0$ is fixed. Indeed, the sum conditions lead to the scheme of generating permutations. Considering that ${\bm p}_k$ is a sequence of the vertices which correspond to the immediate states, it is equivalent to a path in graph theory and we call ${\bm p}_k$ "path index". We can also count the number $n_p(k)$ of ${\bm p}_k$, the number $n_G(k)$ of ${\hat G}_{[{\bm p}_k]}$ and the dimension of $M_{[{\bm p}_k]}$ as follows
\begin{align}
n_p(k)\!=\!{}_{N\!-\!1} P_{k\!-\!1},\hspace{3mm}n_G(k)\!=\!{}_{N\!-\!1} C_{k\!-\!1},\hspace{3mm}{\rm dim}(M_{[{\bm p}_k]})\!=\!N\!\!-\!k.
\end{align}
In calculating $n_G(k)$, we used the fact that ${\hat G}_{[{\bm p}_k]}$ is symmetry on $\tau_1, \tau_2,\ldots\tau_{k-1}$. 

Secondly, we summarize the properties of the recursive Green's function. There are $N-k$ types of the repetition terms $(\alpha\rightarrow \alpha\neq {\tau}_{0},\ldots{\tau}_{k-1})$ in the elements of ${\hat G}_{[{\bm p}_k]}$. In contract, the $k$ types of the repetition terms $(\alpha\rightarrow \alpha={\tau}_{0},\ldots{\tau}_{k-1})$ disappear after $k$ steps. The former is obvious from the definition that ${\hat G}_{[{\bm p}_k]}$ is the propagator on $N-k$ dimensional space $M_{[{\bm p}_k]}$. The reason for the latter is that they are already separated as the diagonal elements of ${\hat G}_{[{\bm p}_i]}$ for $i<k$. In Fig.~\ref{fig:1}, we show the situation of ${\hat G}_{[{\bm p}_k]}$ in this algorithm.
In particular, the type of the repetition terms at $(N-1)$-step is the last one and it means that the iterative scheme has an end. Consequently, we can expand the energy eigenstate $\ket{{\bar \tau}_0}$ by introducing $2^{N-1}$ Green's functions in total.
\begin{figure}[h]
\begin{center}
\includegraphics[width=1.0\linewidth,bb=50 660 305 740]{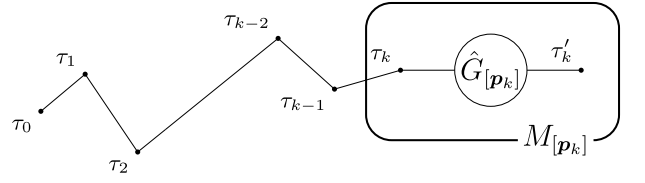}
\caption{\label{1} The recursive Green's function ${\hat G}_{[{\bm p}_k]}$ at $k$-step is introduced via a path ${\bm p}_k$.} 
\end{center}
\end{figure}
\begin{widetext}
Finally, the purpose of the recursive Green's functions is achieved by solving the following recurrence relation whose derivation is shown in Appendix B. 
\begin{equation}
\Ket{g_{\tau_k}}_{[{\bm p}_k]}=\ket{\tau_k}+(-i\lambda)\!\!\!\sum_{\tau_{k+1}\neq{\tau}_{i}\atop i<k+1}^N\!\!\!\Ket{g_{\tau_{k+1}}}_{[{\bm p}_{k+1}]}{G}_{\tau_{k+1}\tau_{k+1}[{\bm p}_{k+1}]}{V_{\tau_{k+1}\tau_{k}}},
\end{equation}
where ${G}_{ij[{\bm p}_{k+1}]}$ and $V_{ij}$ represents the $ij$-th element of ${\hat G}_{[{\bm p}_{k+1}]}$ and ${\hat V}$, respectively. The local transition state on $M_{[{\bm p}_k]}$ which does not contain any repetition terms $(\alpha\rightarrow \alpha={\tau}_{0},\ldots{\tau}_{k})$ is denoted by $\Ket{g_{\tau_k}}_{[{\bm p}_k]}$ as follows
\begin{equation}
\ket{{g}_{\tau_k}}_{[{\bm p}_k]}\equiv\frac{{\hat G}_{[{\bm p}_k]}\ket{\tau_k}}{{\bra {\tau_k}}{\hat G}_{[{\bm p}_k]}{\ket {\tau_k}}}.
\end{equation}
Therefore, considering that $\ket{{g}_{\tau_{N-1}}}_{[{\bm p}_{N-1}]}=\ket{\tau_{N-1}}$ for any ${\bm p}_{N-1}$, we can obtain the $\ket{{g}_{\tau_k}}_{[{\bm p}_k]}$ directly by using the recurrence relation successively
\begin{align}
\ket{g_{\tau_k}}_{[{\bm p}_k]}&=\ket{\tau_k}+(-i\lambda)\!\!\sum_{\tau_{k+1}\neq{\tau}_{i}\atop i<k+1}^N\ket{\tau_{k+1}}{G}_{\tau_{k+1}}{V_{\tau_{k+1}\tau_{k}}}+(-i\lambda)^2\!\!\sum_{\tau_{k+1}\neq{\tau}_{i}\atop i<k+1}^N\!\sum_{\tau_{k+2}\neq{\tau}_{i}\atop i<k+2}^N\ket{\tau_{k+2}}{G}_{\tau_{k+2}}{V_{\tau_{k+1}\tau_{k}}}{G}_{\tau_{k+1}}{V_{\tau_{k+1}\tau_{k}}}\nonumber \\
&\hspace{2mm}+\cdots+(-i\lambda)^{N-k-1}\!\!\sum_{\tau_{k+1}\neq{\tau}_{i}\atop i<k+1}^N\!\sum_{\tau_{k+2}\neq{\tau}_{i}\atop i<k+2}^N\!\cdots\!\!\!\sum_{\tau_{N-1}\neq{\tau}_{i}\atop i<N-1}^N\ket{\tau_{N-1}}\Bigr({G}_{\tau_{N-1}}V_{\tau_{N-1}\tau_{N-2}}\cdots{G}_{\tau_{k+2}}V_{\tau_{k+2}\tau_{k+1}}{G}_{\tau_{k+1}}V_{\tau_{k+1}\tau_{k}}\Bigr).
\end{align}
In writing Eq. (16), we denoted ${G}_{\tau_{k}}\equiv{G}_{\tau_{k}\tau_{k}[{\bm p}_{k}]}$ in order to simplify the expression. Furthermore, in terms of graph theory, the right hand side of Eq. $(16)$ can be also rewritten in the following form
\begin{align}
\ket{\tau_k}+\sum_{{\rm all}\ {\rm paths}\atop{\rm on}\ {M}_{[{\bm p}_{k}]}}\sum_{l=1}^{N-k-1}\!(-i\lambda)^l\ket{\tau_{k+l}}\Bigr({G}_{\tau_{k+l}}V_{\tau_{k+l}\tau_{k+l-1}}\cdots{G}_{\tau_{k+1}}V_{\tau_{k+1}\tau_{k}}\Bigr),
\end{align}
where the sum with respect to the immediate states $\tau_{k+1},\ldots\tau_{k+l}$ runs over all the possible paths on the subspace ${M}_{[{\bm p}_{k}]}$.

Consequently, the energy eigenstate $\ket{\bar{\tau}_{0}}$ can be obtained as a finite series composed of the diagonal elements of the recursive Green's functions and the non diagonal elements of ${\hat V}$
\begin{align}
\ket{\bar{\tau}_{0}}&=\ket{\tau_0}+\sum_{{\rm all}\ {\rm paths}\atop{\rm on}\ M}\sum_{l=1}^{N-1}\!(-i\lambda)^l\ket{\tau_{l}}\Bigr({G}_{\tau_{l}}V_{\tau_{l}\tau_{l-1}}\cdots{G}_{\tau_{1}}V_{\tau_{1}\tau_{0}}\Bigr)\hspace{10mm}(\ z\rightarrow E_{\tau_0}\ )
\end{align}
where we omitted the path index $[{\bm p}_{0}]$ since ${\bm p}_{0}$ is $0$-tuple. Recall that in the representation based on the BW propagator, these diagonal elements ${G}_{\tau_{k}\tau_{k}[{\bm p}_{k}]}$ for $k>0$ represent explicitly all the repetition terms on ${M}_{[{\bm p}_{k}]}$. We need to treat themselves as the effective propagators to drive the Feenberg's perturbation series which has no repetition terms.  Now let us consider the limit $z$ approaches $E_{\tau_0}$ along the positive imaginary axis. This can be formally expressed by replacing $z$ with $E_{\tau_0}+i\epsilon$ where $\epsilon$ is a small positive real number. The expectation value of Eq. (11) in $\ket{\tau_k}$ then obtain the diagonal element ${G}_{\tau_k\tau_k[{\bm p}_k]}$ in the following form of the effective propagator by using Eq. (17)
\begin{align}
{G}_{\tau_k\tau_k[{\bm p}_k]}&=\frac{i}{\bra{\tau_k}(E_{\tau_0}+i\epsilon-{\hat H})\Ket{g_{\tau_k}}_{[{\bm p}_k]}}=\frac{i}{E_{\tau_0}+i\epsilon-\omega_{\tau_k}-\Delta_{\tau_k\tau_k[\bm p_k]}},
\end{align}
where $\Delta_{\tau_k\tau_k[\bm p_k]}$ is a local self-energy defined on $M_{[\bm p_k]}$ and is given by introducing the shifted energies $E_{\tau_k[\bm p_k]}$
\begin{equation}
\hspace{40mm}E_{\tau_k[\bm p_k]}=\omega_{\tau_k}+\Delta_{\tau_k\tau_k[\bm p_k]},\hspace{15mm}k=1,2,\ldots N-1
\end{equation}
\begin{equation}
-i\Delta_{\tau_k\tau_k[\bm p_k]}\equiv(-i\lambda)V_{\tau_k\tau_k}+\!\!\sum_{{\rm all}\ {\rm cycles}\atop{\rm on}\ {M}_{[{\bm p}_{k}]}}\!\biggr(\sum_{l=1}^{N-k-1}(-i\lambda)^{l+1}\frac{V_{\tau_k\tau_{k+l}}\cdots V_{\tau_{k+2}\tau_{k+1}}V_{\tau_{k+1}\tau_k}}{iE_{\tau_{k+l}\tau_0}\cdots iE_{\tau_{k+1}\tau_0}}\!\biggr). \label{eq:wideeq1}
\end{equation}
Similarly, the $\Ket{\bar{\tau}_0}$ can be rewritten
\begin{align}
\Ket{\bar{\tau}_0}&=\Ket{\tau_0}+\sum_{{\rm all}\ {\rm paths}\atop{\rm on}\ M}\biggr(\sum_{l=1}^{N-1}(-i\lambda)^{l}\frac{V_{\tau_{l}\tau_{l-1}}\cdots V_{\tau_{2}\tau_{1}}V_{\tau_{1}\tau_{0}}}{iE_{\tau_{l}\tau_{0}}\cdots iE_{\tau_{2}\tau_{0}}iE_{\tau_{1}\tau_{0}}}\Ket{\tau_{l}}\biggr).
\end{align}
In writing Eq. (21), we denoted $E_{\tau_i\tau_j}=E_{\tau_i[\bm p_i]}-E_{\tau_j[\bm p_j]}$ and $E_{\tau_0[\bm p_0]}=E_{\tau_0}+i\epsilon$ and the sum with respect to the immediate states $\tau_{k+1},\ldots\tau_{k+l}$ runs over all the possible cycles on the subspace $M_{[\bm p_k]}$. It is important that the above energy eigenstate is composed of the finite paths, while the local self energies are composed of the finite cycles. Therefore, as long as $N$ is finite, these are the exact equations without truncating the diagrams perturbatively.
On the other hands, the representation of the eigenstate in Eq. (22) obviously require the unknown exact energy $E_{\tau_0}$ for its calculation. In the next subsection, we will drive an exact self-consistent equation to determine the energy.
\end{widetext}
\subsection{Exact self-consistent equation}
First of all, in order to drive an exact self-consistent equation for the exact energy $E_{\tau_0}$,  we should note that all the shifted energies can be expressed as the functions of $E_{\tau_{0}[\bm p_0]}$ composed of the finite continued fractions. This can be shown in the following manner. 

Eq. (20) is a system of $n_{E}(N)$ equations in the shifted energies $E_{\tau_k[\bm p_k]}$, $n_{E}(N)$ is the total number of $E_{\tau_k[\bm p_k]}$ from $k=1$ to $N-1$
\begin{align}
n_{E}(N)=\sum_{k=1}^{N-1}(N-k)n_G(k)=(N-1)2^{N-2},
\end{align}where we used the fact that the number of $E_{\tau_k[\bm p_k]}$ for fixed $k>0$ is equal to that of $G_{\tau_k\tau_k[\bm p_k]}$ which is $(N-k)n_G(k)$, and gives the recursive relation in $E_{\tau_k[\bm p_k]}$ for $k>0$
\begin{align}
E_{\tau_k[\bm p_k]}\!=\!\omega_{\tau_k}\!\!+\!\Delta_{\tau_k\tau_k[\bm p_k]}(\!E_{\tau_{k+1}\![\bm p_{k+1\!}]}\!\ldots\!E_{\tau_{N-1}\![\bm p_{N-1\!}]};\!E_{\tau_{0}[\bm p_0]}).
\end{align}
Considering that $E_{\tau_{N-1}[\bm p_{N-1}]}=\omega_{\tau_{N-1}}+\lambda V_{\tau_{N-1}\tau_{N-1}}$ for any ${\bm p}_{N-1}$, we can formally express $E_{\tau_k[\bm p_k]}$ as a function of $E_{\tau_0}$ by using Eq. (24) from $N-1$ to $k$ successively
\begin{equation}
E_{\tau_k[\bm p_k]}=Z_{\tau_k[\bm p_k]}(E_{\tau_{0}[\bm p_0]}),\hspace{3mm}k=1, 2,\ldots N-1
\end{equation}
where $Z_{\tau_k[\bm p_k]}$ is composed of the finite continued fractions.

Next, the above equation suggests that there are not enough equations to determine the exact energy $E_{\tau_0}$. In fact, the equation for $E_{\tau_0}$ is derived separately from Eq. (20) in our approach and the pole condition that ${G}_{\tau_0\tau_0}^{-1}$ vanishes as $\epsilon$ approaches 0, which is a necessary and sufficient condition for the assumption $\Braket{\tau_0|\psi_{\tau_0}}\neq0$, gives the following equation in the same representation as Eq. (20) when $k=0$
\begin{align}
E_{\tau_{0}[\bm p_0]}=\omega_{\tau_0}+\Delta_{\tau_0\tau_0}(E_{\tau_{1}[\bm p_{1}]}\ldots E_{\tau_{N-1}[\bm p_{N-1}]}; E_{\tau_{0}[\bm p_0]}).
\end{align}
Then the exact self-consistent equation for $E_{\tau_0}$ can be obtained in the form of the finite continued fractions by eliminating all the shifted energies from Eq. (26) with Eq. (25) 
\begin{align}
E_{\tau_{0}[\bm p_0]}=Z_{\tau_0}(E_{\tau_{0}[\bm p_0]}),\hspace{4mm}(\epsilon\rightarrow0)
\end{align}
Finally, we can calculate $E_{\tau_0}$ with arbitrary accuracy by using the numerical methods such as an iterative and Steffensen's method\cite{12} and thus the eigenstate $\Ket{\bar{\tau}_0}$ is determined by obtaining all the shifted energies from Eq. (25).
\section{\label{sec:level3}ANALYSIS OF EFFECTIVE PROPAGATORS}
In this section, we analyze the diagonal elements of the recursive Green's functions in terms of the Neumann series and determinant representation for them. 
\subsection{Resummation}
There is an interesting relationship between the Rayleigh-Schr${\ddot{\rm o}}$dinger perturbation and Brillouin-Wigner perturbation theories which was mentioned by Lennard-Jones: the RS series is driven by expanding the unperturbed propagators in the BWPT with respect to a small self energy\cite{5,13}. In other wards, the BW series can be regarded as a resummation of the RS series. Furthermore, the Feenberg's perturbation series is no exception and closely related to these perturbation theories.

\begin{figure}[!h]
\begin{center}
\includegraphics[width=1.0\linewidth,bb=50 680 305 740]{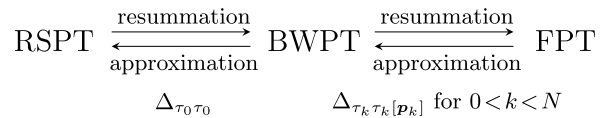}
\caption{\label{fig:2}The relationships between the Rayleigh-Schr${\ddot{\rm o}}$dinger(RSPT), Brillouin-Wigner(BWPT) and Feenberg(FPT) perturbation theories. The resummation and approximation with respect to the self energies play a role in connecting between the three theories.} 
\end{center}
\end{figure}

In the FPT derived from the recursive Green's function approach, their diagonal elements have also two representations of the perturbation series based on the BW propagator $G_{BW}$ and effective propagator ${G}_{\tau_k\tau_k[{\bm p}_k]}$ in the FPT. Indeed, a geometric series expansion for ${G}_{\tau_k\tau_k[{\bm p}_k]}$ with respect to the local self-energy $\Delta_{\tau_k\tau_k[\bm p_k]}$ leads to the following expression
\begin{align}
&G_{BW}+G_{BW}(-i\Delta_{\tau_k\tau_k[\bm p_k]})G_{BW}\nonumber\\
&\hspace{5mm}+G_{BW}(-i\Delta_{\tau_k\tau_k[\bm p_k]})G_{BW}(-i\Delta_{\tau_k\tau_k[\bm p_k]})G_{BW}+\cdots
\end{align}
where
\begin{align}
G_{BW}\equiv\frac{i}{E_{\tau_0}-\omega_{\tau_k}},\hspace{7mm}\Delta_{\tau_k\tau_k[\bm p_k]}=\mathcal{O}(\lambda).
\end{align}
Note that the estimation for the local self-energy can be justified only in non-degenerate systems (See Appendix C). 
Eq. (28) means that the effective propagator can be expressed as an infinite series consisting of the BW propagators and local self-energies while all the repetition terms $(\tau_{k} \rightarrow \tau_{k})$ on $M_{[\bm p_k]}$ can be expressed as a chain of the cycles composed of the effective propagators in terms of diagrams. Furthermore, the expansion for $\Delta_{\tau_k\tau_k[\bm p_k]}$ itself causes the production of a huge number of another cycles $(i \rightarrow i\neq \tau_{k},\ldots\tau_0)$ because they have a nested structure of themselves. In this way, all the repetition terms are exposed as an infinite series based on the BW propagator. Therefore we can interpret that the BWPT is an approximation to the FPT, or conversely, the FPT is the resummed perturbation theory by modifying representations of their propagators.  Now we show the relation between the FPT, BWPT and RSPT in Fig.~\ref{fig:2}. 
\subsection{Rearrangement of the path-indices}
\begin{figure}[h]
\begin{center}
\includegraphics[width=1.0\linewidth,bb=55 640 300 750]{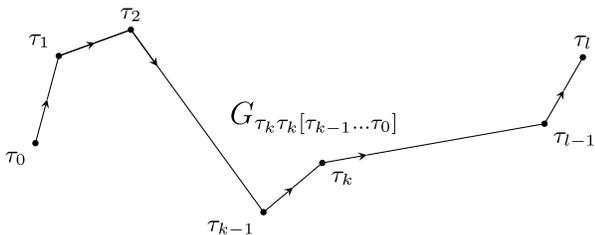}
\caption{\label{fig:3} The subscripts $\tau_{k-1}\ldots\tau_0$ of our effective propagator ${G}_{\tau_k\tau_k[\tau_{k-1}\ldots\tau_0]}$ at $k$-step indicate that the propagator is connected to a subgraph ($\tau_0\rightarrow\tau_1\cdots\rightarrow\tau_{k-1}$)} 
\end{center}
\end{figure}

We should note that the Feenberg's perturbation series shown in previous section is different from the original one proposed by Feenberg in terms of the arrangement of the effective propagators and intermediate states dependency. In order to clarify the differences, we consider a general term in the eigenstate which corresponds to the transition process ($\tau_0\rightarrow\tau_1\ldots\rightarrow\tau_l$) shown in Fig.~\ref{fig:3}.  

Firstly, it is obvious from Eq. (22) that our approach gives a finite product of the effective propagators included in the general term 
\begin{align}
{G}_{\tau_l\tau_l[\tau_{l-1}\ldots\tau_0]}\cdots{G}_{\tau_{2}\tau_{2}[\tau_1\tau_0]}{G}_{\tau_1\tau_1[\tau_0]}.
\end{align}
The effective propagator ${G}_{\tau_k\tau_k[\tau_{k-1}\ldots\tau_0]}$ for $k\leq l$ depends on $k$ intermediate states from $\tau_0$ to $\tau_{k-1}$ and thus that can provide an interpretation of its path-index: the effective propagator with $\bm p_k$ is connected to the path $\bm p_k$ which is a subgraph in the transition process.
On the other hands, that of the effective propagators in his paper is given by using the expression of the recursive Green's function as follows 
\begin{align}
{G}_{\tau_l\tau_l[\tau_0]}\cdots{G}_{\tau_{2}\tau_{2}[\tau_{3}\dots\tau_{l}\tau_0]}{G}_{\tau_1\tau_1[\tau_{2}\ldots\tau_{l}\tau_0]},
\end{align}
where ${G}_{\tau_{k}\tau_{k}[\tau_{k+1}\dots\tau_{l}\tau_0]}$ at $k$-step is formally obtained by replacing the subscripts $\tau_{l-k+1}\ldots\tau_0$ of ${G}_{\tau_{l-k+1}\tau_{l-k+1}[\tau_{l-k}\ldots\tau_0]}$ with those $\tau_{k}\dots\tau_{l}\tau_0$, respectively. Thus the number of the intermediate states on which his effective propagator at $k$-step depends is $l-k+1$ unlike our approach. These differences arise mainly from some recursive procedures to derive the FPT. For example, the projection operator method has a same arrangement as our approach while the latter is employed in the continued fraction methods and Feshbach's successive approximation. Thus, despite the same transition process, the equations for calculating the transition amplitude are different in appearance. 

In fact, both factors are correct and can be transformed into each other in the following manner. Let ${\bm H}$ denote the matrix representation of the full Hamiltonian $\hat H$ in terms of the unperturbed eigenstate. Then the determinant representation for the effective propagator is given as follows
\begin{align}
-i{G}_{\tau_k\tau_k[{\bm p}_k]}=\frac{{\ \ \bigr|E_{\tau_0}+i\epsilon-{\bm H}_{[{\bm p}_{k+1}]}\bigr|}}{{\bigr|E_{\tau_0}+i\epsilon-{\bm H}_{[{\bm p}_k]}\bigr|}},
\end{align}
where ${\bm H}_{[{\bm p}_k]}$ is the $(N-k)\times(N-k)$ submatrix obtained by deleting the ${\tau}_{0},{\tau}_{1},\ldots{\tau}_{k-1}$-th rows and columns of ${\bm H}$. This representation is derived from Eq. (11) by using the Cramer's rule (See Appendix D) and is equivalent to the expression of the effective propagator in the continued fraction method\cite{9}. Applying Eq. (32) to both Eqs. (30) and (31), we  can show that they have the same results as follows
\begin{align}
i^l\frac{{\ \ \bigr|E_{\tau_0}+i\epsilon-{\bm H}_{[\tau_l\ldots\tau_0]}\bigr|}}{{\bigr|E_{\tau_0}+i\epsilon-{\bm H}_{[\tau_0]}\bigr|}},
\end{align}
where we canceled the common factors in the numerator and denominator.

In this way, it is proved that the two expressions for the general term are equivalent in terms of the determinant representation. In general, we can also show that there are $l!$ expressions of the term for arbitrary rearrangement of the effective propagators from the above calculation procedure.
\subsection{Removable and non removable singularities}
In order to complete this section, we briefly discuss the pole of the effective propagator. In the RSPT and BWPT, if an energy denominator which appears in the perturbation series becomes zero or nearly zero, then the perturbation theories break down due to the divergence of the propagator. This often occurs in the degenerate case and hence the formulations for degenerate systems has been developed to avoid that.

In the Feenberg's perturbation series, there is a possibility that the effective propagator diverges in the limit $z\rightarrow E_{\tau_0}$. For non-degenerate systems, we can show that all the energy denominators have a finite gap $E_{\tau_i[\bm p_i]}-E_{\tau_0[\bm p_0]}\approx\omega_{\tau_i}-\omega_{\tau_0}$ (See Appendix C). All the effective propagators thus have no poles at $z=E_{\tau_0}$ in this case. On the other hands, it is not clear for degenerate systems whether the effective propagator has no poles at $z=E_{\tau_0}$ because the above manner does not work here. 

Now we consider the poles of the effective propagator ${G}_{\tau_k\tau_k[{\bm p}_k]}$ for $1\leq k<N$ to clarify the specific conditions in which the FPT fails. From Eq. (19) the zeros of  ${G}^{-1}_{\tau_k\tau_k[{\bm p}_k]}$ gives the conditions as a local self-consistent equation on the subspace $M_{[\bm p_k]}$ formally 
\begin{align}
E_{\tau_0[\bm p_0]}=Z_{\tau_k[\bm p_k]}(E_{\tau_{0}[\bm p_0]}),
\end{align}
which are also represented from Eq. (32) as a secular equation
\begin{align}
{\bigr|E_{\tau_0}+i\epsilon-{\bm H}_{[\bm p_k]}\bigr|}=0.
\end{align}
These mean that if an eigenvalue $E_{\tau_0}$ of ${\bm H}$ is equal to that of the submatrix ${\bm H}_{[\bm p_k]}$, then the effective propagator diverges. We thus reach a tentative conclusion that there are $n_{E}(N)$ singularities, counted with multiplicity, in the Feenberg's perturbation series. Recall, however, that the Feenberg's formulations have a continued fraction structure for $E_{\tau_0}$.  Most of the singularities are classified as the removable singularities in terms of cancellation of pole and zero. Indeed all poles given by Eq. (35) except $k=0$ disappear from the general terms Eq. (33) while there is a determinant ${\bigr|E_{\tau_0}+i\epsilon-{\bm H}_{[\tau_0]}\bigr|}$ in the denominator. 

From the above analysis of the pole structure of the effective propagators, it became clear that the Feenberg's perturbation series has at most $N-1$ poles which are given as the zeros of ${\bigr|E_{\tau_0}+i\epsilon-{\bm H}_{[\tau_0]}\bigr|}$. These poles also occur in the self-consistent equation for $E_{\tau_0}$ and thus their existence is an important factor to explore the validity of the FPT. In Sec.~\ref{sec:level2}, we have restricted ourselves to the non-degenerate case in order to ensure that the assumption $\Braket{\tau_0|\psi_{\tau_0}}\neq0$ is satisfied. This is suitable to confirm the resummation mechanism of the effective propagators, while a tight restriction in comparison to the original assumption. Now we can in fact relax some restrictions by imposing the following regularity condition on the eigenvalue $E_{\tau_0}$ instead of the assumption
\begin{equation}
{\bigr|E_{\tau_0}+i\epsilon-{\bm H}_{[\tau_0]}\bigr|}\neq0,\hspace{5mm}(\ \epsilon\rightarrow 0\ ).
\end{equation}
Note that this is a sufficient condition for the assumption and its sufficiency is proved in Appendix A.
It is important that the Feenberg's formulations are always available regardless of degeneracy and regardless of the magnitude of $\lambda$ as long as the condition Eq. (36) is satisfied.
\section{\label{sec:level4}NUMERICAL CALCULATION}
\subsection{Convergences}
The self-consistent equation for the energy in the Feenberg perturbation theory is exact and we can obtain it with arbitrary accuracy by solving the equation numerically. It is supposed that this high convergence is caused by incorporating all the repetition terms in the resummed form of Eq. (28). In this subsection, we compare the convergence speed between the FPT, BWPT and RSPT. Now we propose the following approximation to the FPT
\begin{equation}
E_{\tau_k[\bm p_k]}=\omega_{\tau_k}+\Delta^{(m)}_{\tau_k\tau_k[\bm p_k]}+{\mathcal O}(\lambda^{m+1}),\hspace{3mm}k=0, 1,\ldots N-1
\end{equation}
where $\Delta^{(m)}_{\tau_k\tau_k[\bm p_k]}$ is the local self energy up to $m$-th order of $\lambda$.
This approximation is based on the interpretation that the local self-energies play a role of the fundamental corrections to the energy denominators in the FPT. 
\begin{widetext}
The self-consistent equation given by Eq. (37) in the same way as Eq. (27) contains the local self energies up to $m$-th order exactly. Therefore, considering the effect of resummation, we can expect that the $m$-th order FPT often gives the high precision approximate solutions compared to $m$-th order BWPT and RSPT. Let ${\bm H_0}$ and ${\bm V}$ denote the matrix representation of the unperturbed Hamiltonian ${\hat H}_0$ and interaction term $\hat V$ in terms of the unperturbed eigenstate, respectively and we consider the following example in the case $N=7$ in order to check the accuracy of the FPT
\begin{equation}
{\bm H_0}={\rm diag}(\omega_1, \omega_2, \omega_3, \omega_4, \omega_5, \omega_6, \omega_7)={\rm diag}(0.07, 0.30, 0.37, 0.48, 0.51, 0.80, 0.86), \nonumber
\end{equation}
\begin{equation}
{\bm V}\!=\!
\begin{bmatrix}
(0.26, \ \ \pi\ ) &(0.09, 4.48) &(0.38, 3.39) &(0.66, 4.82) &(0.97, 2.66) &(0.37, 3.97) &(0.05, 0.53)\\
             &  (0.15, \ \ \pi\ ) &(0.88, 0.18) &(0.95, 3.57) &(0.49, 1.10) &(0.09, 5.68) &(0.70, 1.95)\\
            &                &(0.48, \ \ \pi\ )   &(0.01, 1.47) &(0.43, 1.04) &(0.59, 0.77) &(0.72, 6.06)\\
		   &			     &			  &(0.44, \ \ \pi\ )   &(0.68, 5.91) &(0.17, 1.04) &(0.52, 4.69)\\
		   &			     &			  &			    &(0.49, 0.00) &(0.99, 3.63) &(0.96, 3.40)\\
		   &  \hsymb{*}&		       &			    &			 &(0.46, 0.00) &(0.79, 6.26)\\
		   &                &                &                &                 &                &(0.22, \ \ \pi\ )\nonumber
\end{bmatrix},
\end{equation}
where $V_{ij}=V_{ji}^{*}$ for $j<i$. The elements of ${\bm H_0}$ and ${\bm V}$ are randomly selected from the ranges $0\leq \omega_i\leq1$ and $|V_{ij}|\leq1$, respectively and we denote the $V_{ij}$ as a point $(|V_{ij}|,\ {\rm arg}(V_{ij}))$ for $0\!\leq\!{\rm arg}(V_{ij})\!<\!2\pi$ in polar coordinates. In Fig.~\ref{fig:4}, we fix a coupling constant $\lambda=0.01$ and plot the geometric mean of relative errors (GMRE) as a function of perturbation order $m$ which is defined as follows
\begin{equation}
{\rm GMRE}(m)\equiv\sqrt[7]{\prod_{i=1}^{7}\ \biggr{|}\frac{E_i^{(m)}-E_i^{\rm{exact}}}{E_i^{\rm{exact}}}\biggr{|}},
\end{equation}
where $E_i^{(m)}$ is the $i$-th energy eigenvalue given by each $m$-th order perturbation theory and we calculated them by Steffensen's method. 
\begin{figure}[!h]
\begin{center}
\includegraphics[width=0.62\linewidth]{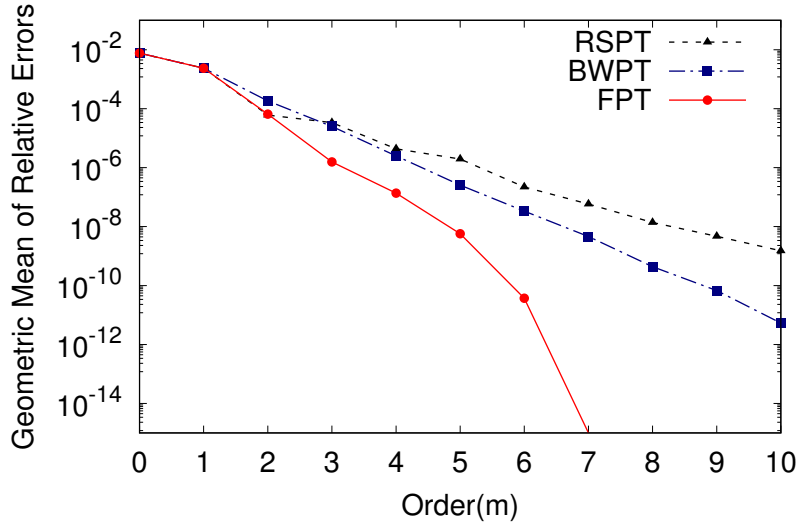}
\end{center}
\caption{\label{fig:4}Comparison of geometric means of relative errors between the Rayleigh-Schr${\ddot{\rm o}}$dinger(RSPT), Brillouin-Wigner(BWPT) and Feenberg(FPT) perturbation theories when $\lambda=0.01$}
\end{figure}

Note that the three theories give the same result at zero and first order because there is no difference in their formulas. For second or above order, we find that the convergence of the FPT is equal to or more rapid than that of the BWPT and RSPT on average. These results show the relationships between the three theories in Fig.~\ref{fig:2} and support the acceleration convergence due to the effect of resummation. In particular, the $7$-th order FPT returns to the exact theory and can give all the energy eigenvalues with a relative error less than $10^{-15}$. Consequently, the MGRE($7$) is also less than $10^{-15}$ as shown in Fig.~\ref{fig:4}.
\end{widetext}

\subsection{Strong coupling region}
To verify the validity of the FPT, we consider again the previous example and calculate the energy eigenvalues and eigenstates of ${\hat H}_0+\lambda{\hat V}$ when the value of $\lambda$ is increased from $0$ to $1$. 
In this calculation, we assume that the above sufficient condition is satisfied and regard the exact self-consistent equation for $E_{\tau_0}$ as a non linear equation with a free parameter $\lambda$
\begin{align}
E_{\tau_0}=Z_{\tau_0}(E_{\tau_0},\lambda).
\end{align}
Putting an initial value $E_{\tau_0}=\omega_{\tau_0}$ at $\lambda=0$ and we traced the solution curve which is called "eigenpath" one by one. Furthermore, in order to avoid the jump to other eigenpaths, we used the spherical algorithm\cite{14} and also employed the Steffensen's method as an iterative scheme in terms of $E_{\tau_0}$.

\begin{figure}[!t]
\begin{center}
\subfloat[The growth process of the energy eigenvalues.]{\includegraphics[width=1.0\linewidth]{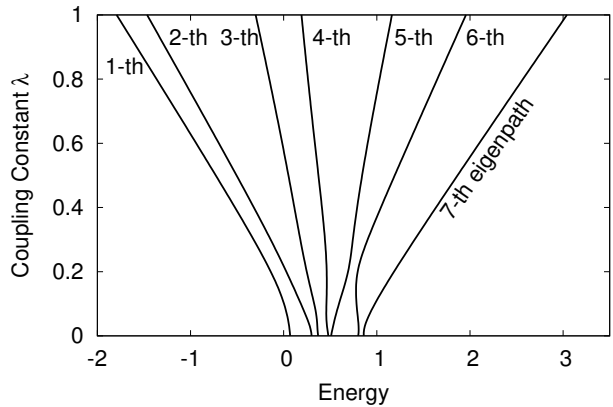}\label{fig:5a}}\\
\subfloat[The probabilities of observing the unperturbed eigenstates which are contained in the ground state $\ket{\bar{1}}$.]{\includegraphics[width=1.0\linewidth]{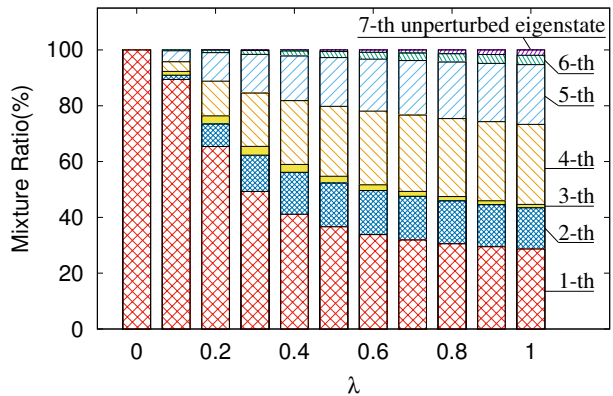}\label{fig:5b}}
\end{center}
\caption{The behavior of the eigenvalues and eigenstates in strong coupling region}
\end{figure}

Fig.~\ref{fig:5a} shows the growth process of the eigenvalues beyond the perturbative region ($\lambda\ll1$). In general, the RSPT and BWPT only work in the weak coupling region and cannot trace the eigenvalues to the strong one. Therefore this construction of the eigenpaths is one of the advantages of the FPT. Furthermore, we notice that each eigenpath spreads in a fan shape as the $\lambda$ increases. This comes from the fact that each eigenvalue asymptotically approaches the one of $\lambda{\hat V}$ which is a dominant factor in the strong coupling region.

In Fig.~\ref{fig:5b}, we illustrate the composition ratio of the ground state $\ket{\bar{1}}$ as a probability of observing the unperturbed eigenstates. The unperturbed component decreases with $\lambda$ while the other perturbed components increase. In particular, the large mixing between the unperturbed eigenstates is caused in $\lambda\sim1$ and it is obviously non perturbative state. It can thus be said that the FPT is available in the strong coupling region.

\section{\label{sec:level5}SUMMARY}
In this paper, using the recursive Green's function approach we have developed the new derivation method of the Feenberg perturbation theory for $N$-level quantum systems. The main advantages of our approach can be summarized in the following two points.
\begin{itemize}
\item The Feenberg's perturbation series is directly derived from a simple recurrence relation in Eq. $(14)$ as a natural generalization of the Brillouin-Wigner perturbation series.
\item  There are three aspects in the diagonal element of the recursive Green's functions: the effective propagators in the FPT, all the repetition terms on a subspace and determinant representation in Eq. $(32)$.
\end{itemize}
Our recurrence relation is based on the extraction of the perturbed eigenstate in Eq. $(4)$ and generalization of scheme of removing the repetition terms in the BWPT. Then all the repetition terms are separated as the diagonal elements of the recursive Green's functions systematically and thus our approach has a simpler recursive structure then the other approaches. 

Furthermore we can easily reproduce the formulations given by another approaches\cite{8,9,10,11} by using the determinant representation or rearranging the order of the effective propagators in the perturbation series. This is one of the most important results in this paper. 
\appendix
\section{PROOF OF EQ. (4) AND EQ. (36)}
\subsection{Proof of Eq. (4)}
In this appendix, we prove that if the ${\tau_0}$-th element of the eigenstate $\ket{\psi_{\tau_0}}$ has a non-zero value, then the limit given in Eq. (4) is an unnormalized eigenstate $\ket{\bar{\tau}_{0}}$.

First of all, for simplicity, we assume that the spectrum of $\hat H$ is not degenerate: $E_i\neq E_j$ for $i\neq j$. Using the spectral representation Eq. (3), the numerator and the denominator of the left-hand side of Eq. (4) can be rewritten respectively as  
\begin{align}
{\hat G}(z)\ket{\tau_0}&=\sum_{\alpha=1}^{N}\ket{\psi_\alpha}\frac{iC^*_{\alpha}}{z-E_\alpha},\\
\bra{\tau_0}{\hat G}(z)\ket{\tau_0}&=\sum_{\alpha=1}^{N}\frac{i|C_{\alpha}|^2}{z-E_\alpha}.
\end{align}
Note that $C_{\alpha}$ denotes the ${\tau_0}$-th element $\Braket{\tau_0|\psi_{\alpha}}$ of the $\alpha$-th eigenstate and $C_{\tau_0}\neq0$ on the assumption. It is thus clear that both the numerator and the denominator have a simple pole at $z=E_{\tau_0}$. 

Next, as $z\rightarrow E_{\tau_0}$, we can evaluate the dominant terms in the numerator and the denominator as follows
\begin{align}
{\hat G}(z)\ket{\tau_0}\sim\ket{\psi_{\tau_0}}\frac{iC^*_{\tau_0}}{z-E_{\tau_0}},\hspace{4mm}\bra{\tau_0}{\hat G}(z)\ket{\tau_0}\sim\frac{i|C_{\tau_0}|^2}{z-E_{\tau_0}},
\end{align}
and then the limit in Eq. (4) has a finite result because its pole and zero at $z=E_{\tau_0}$ cancel each other
\begin{equation}
\lim_{z \to E_{\tau_0}}\frac{{\hat G}(z)\ket{\tau_0}}{\bra{\tau_0}{\hat G}(z)\ket{\tau_0}}=\frac{1}{C_{\tau_0}}\ket{\psi_{\tau_0}}.
\end{equation}
This result is certainly proportional to the $\tau_0$-th perturbed eigenstate with constant of proportionality $C^{-1}_{\tau_0}$ and we can obtain Eq.(4) by redefining the right hand-side of Eq. (A4) as the unnormalized eigenstate $\ket{\bar{\tau}_{0}}$.

Note also that this extraction of the eigenstate works in the degenerate case. Now we assume that $\hat H$ has $d$ degenerate eigenstates $\ket{\psi_{\beta_1\equiv\tau_0}}, \ket{\psi_{\beta_2}},\ldots\ket{\psi_{\beta_d}}$. By following the same procedure as the above we can indeed obtain the following limit  
\begin{equation}
\lim_{z \to E_{\tau_0}}\frac{{\hat G}(z)\ket{\tau_0}}{\bra{\tau_0}{\hat G}(z)\ket{\tau_0}}=\sum_{\alpha\in D}{\theta}_{\alpha}\ket{\psi_{\alpha}},
\end{equation}
where $D$ is a set $\{\beta_1\equiv\tau_0, \beta_2,\ldots \beta_d\}$ and $\theta_\alpha$ is given as follows
\begin{equation}
{\theta}_{\alpha}=\frac{C^*_{\alpha}}{\sum_{\alpha'\in D}|C_{\alpha'}|^2}.
\end{equation}
Furthermore we have the following assumption which occurs as a result of the extensions
\begin{equation}
\sum_{\alpha\in D}|C_{\alpha}|^2\neq0.
\end{equation}
In Eq. (A6), although the right hand-side is expressed as a superposition of the degenerate eigenstates, it is certainly the eigenstate of $\hat H$. Therefore we can conclude that Eq. (4) is always satisfied on the assumptions in both degenerate and non-degenerate cases.
\subsection{Proof of Eq. (36)}
To prove the sufficiency of the regularity condition Eq. (36), we start with the two representation for ${G}_{\tau_0\tau_0}$. 

One is the spectral representation in Eq. (A2) and it can be seen that ${G}_{\tau_0\tau_0}$ has a simple pole at $z=E_{\tau_0}$ and its residue is $C_{\tau_0}=\Braket{\tau_0|\psi_{\alpha}}\neq0$. Then we can give the following equation for $E_{\tau_0}$ from the relation between poles and zeros of ${G}_{\tau_0\tau_0}$  
\begin{equation}
\lim_{z \to E_{\tau_0}}{G}^{-1}_{\tau_0\tau_0}=0,
\end{equation}
Note that this is still a necessary and sufficient condition for the assumption $\Braket{\tau_0|\psi_{\alpha}}\neq0$. Using the determinant representation, Eq. (A8) can be rewritten as
\begin{equation}
\lim_{\epsilon \to 0}i\frac{{\bigr|E_{\tau_0}+i\epsilon-{\bm H}\bigr|}}{{\bigr|E_{\tau_0}+i\epsilon-{\bm H}_{[\tau_0]}\bigr|}}=0.
\end{equation}
Thus it is clear that the regularity condition $\bigr|E_{\tau_0}+i\epsilon-{\bm H}_{[\tau_0]}\bigr|\neq0$ is the sufficient condition for Eq. (A9). This completes the proof of the sufficiency.
\section{DERIVATION OF RECURRENCE RELATION}
The recurrence relation Eq. (14) is driven from the definition Eq. (11). First, considering that ${\hat G}_{[{\bm p}_k]}$ does not have $\tau_0\ldots\tau_{k-1}$-th rows and columns, we focus on $\tau_k$-th column vector of Eq. (11) such as $\tau_k\neq\tau_0\ldots\tau_{k-1}$ 
\begin{align}
{\hat P}_{[{\bm p}_k]}(z-{\hat H}){\hat P}_{[{\bm p}_k]}{\hat G}_{[{\bm p}_k]}{\ket {\tau_k}}&=i{\ket {\tau_k}}.
\end{align}
Applying ${\hat P}_{[{\bm p}_{k+1}]}$ or ${\Bra{\tau_k}}$ to Eq. (B1) from the left side and using the division ${\hat P}_{[{\bm p}_k]}={\hat P}_{[{\bm p}_{k+1}]}+\Ket{\tau_k}\!\Bra{\tau_k}$, we obtain the two following equations 
\begin{align}
{\hat P}_{[{\bm p}_{k+1}]}(z-{\hat H})({\hat P}_{[{\bm p}_{k+1}]}+\Ket{\tau_k}\!\Bra{\tau_k}){\hat G}_{[{\bm p}_k]}{\ket {\tau_k}}&=0,\\
{\Bra{\tau_k}}(z-{\hat H})({\hat P}_{[{\bm p}_{k+1}]}+\Ket{\tau_k}\!\Bra{\tau_k}){\hat G}_{[{\bm p}_k]}{\ket {\tau_k}}&=i,
\end{align}
where we used ${\hat P}_{[{\bm p}_{k+1}]}{\hat P}_{[{\bm p}_{k}]}={\hat P}_{[{\bm p}_{k+1}]}$. These are equivalent to Eq .(B1) in all. In particular, Eq. (B2) leads to the recurrence relation Eq. (11), while we should note that Eq. (B3) yields Eq. (19) which is associated with the diagonal elements ${G}_{\tau_k\tau_k[{\bm p}_k]}$. 

Next, we define the local transition state $\ket{{g}_{\tau_k}}_{[{\bm p}_k]}$ on the subspace $M_{[\bm p_k]}$.
\begin{equation}
\ket{{g}_{\tau_k}}_{[{\bm p}_k]}\equiv\frac{{\hat G}_{[{\bm p}_k]}\ket{\tau_k}}{{\bra {\tau_k}}{\hat G}_{[{\bm p}_k]}{\ket {\tau_k}}}={\ket {\tau_k}}+\ket{g'_{\tau_k}}_{[{\bm p}_k]},
\end{equation}
where $\ket{g'_{\tau_k}}_{[{\bm p}_k]}$ is an perturbed part ${\hat P}_{[{\bm p}_{k+1}]}\ket{{g}_{\tau_k}}_{[{\bm p}_k]}$. Eq. (B2) can then be rewritten as a relation between the perturbed part $\ket{g_{\tau_k}}_{[{\bm p}_k]}$ and unperturbed part $\Ket{\tau_k}$ in $M_{[\bm p_k]}$.
\begin{align}
{\hat P}_{[{\bm p}_{k+1}]}(z-{\hat H}){\hat P}_{[{\bm p}_{k+1}]}\ket{g'_{\tau_k}}_{[{\bm p}_k]}&=\lambda{\hat P}_{[{\bm p}_{k+1}]}\hat{V}\Ket{\tau_k}.
\end{align}
This is a linear equation for $\ket{g'_{\tau_k}}_{[{\bm p}_k]}$ and by introducing the recursive Green's function ${\hat G}_{[{\bm p}_{k+1}]}$ at ($k+1$)-step, we can formally obtain the $\ket{g'_{\tau_k}}_{[{\bm p}_k]}$ as follows
\begin{align}
\Ket{g'_{\tau_k}}_{[{\bm p}_k]}&=\lambda{\hat G}_{[{\bm p}_{k+1}]}{\hat P}_{[{\bm p}_{k+1}]}\hat{V}\Ket{\tau_k}.
\end{align}
Finally, taking back Eq. (B6) to Eq. (B4) and rearranging it, we obtain the recurrence relation Eq. (14)
\begin{equation}
\Ket{g_{\tau_k}}_{[{\bm p}_k]}\!\!=\!\ket{\tau_k}+(-i\lambda)\!\!\!\!\!\sum_{\tau_{k+1}\neq{\tau}_{i}\atop i<k+1}^N\!\!\!\!\!\Ket{g_{\tau_{k+1}}}_{\![{\bm p}_{k+1}\!]}\!{G}_{\tau_{k+\!1}\!\tau_{k+\!1}[{\bm p}_{k+\!1}\!]}\!{V_{\tau_{k+\!1}\!\tau_{k}}}.
\end{equation}
\section{EVALUATION OF THE LOWEST ORDER LOCAL SELF-ENERGIES AND SHIFTED ENERGIES}
Now, we start with zero order approximation for an energy eigenvalue $E_{\tau_0}$ and $E_{\tau_{N-1}[\bm p_{N-1}]}\!=\!\omega_{\tau_{N-1}}\!+\!\lambda V_{\tau_{N-1}\tau_{N-1}}$
\begin{align}
E_{\tau_0}\approx\omega_{\tau_0},\hspace{5mm}E_{\tau_{N-1}[\bm p_{N-1}]}\approx\omega_{\tau_{N-1}}.
\end{align}
Hence, in the non-degenerate case, the energy differences $E_{\tau_{N-1}[\bm p_{N-1}]}-E_{\tau_0}$ at ($N-1$)-step have a finite gap $\omega_{\tau_{N-1}}-\omega_{\tau_0}(\tau_{N-1}\neq\tau_0)$. 
Then we can evaluate the local self-energy and shifted energy at $(N-2)$-step from Eqs. (20)(21)
\begin{align}
\Delta_{\tau_{N-2}\tau_{N-2}[\bm p_{N-2}]}=\mathcal{O}(\lambda),\hspace{3mm}E_{\tau_{N-2}[\bm p_{N-2}]}\approx\omega_{\tau_{N-2}}.
\end{align}
Using Eqs. (20)(21) from $k=N-2$ to $1$ successively, all the local self-energies and shifted energies can be evaluated as follows
\begin{align}
\Delta_{\tau_{k}\tau_{k}[\bm p_{k}]}=\mathcal{O}(\lambda),\hspace{5mm}E_{\tau_{k}[\bm p_{k}]}\approx\omega_{\tau_{k}}
\end{align}
where $k=1,2,\ldots N-1$.
\section{DETERMINANT REPRESENTATION}
Application of Cramer's rule in Eq. (11) gives the determinant representation for the effective propagator ${G}_{\tau_k\tau_k[{\bm p}_k]}$
\begin{align}
(z-{\bm H}_{[{\bm p}_k]}){\bm G}_{[{\bm p}_k]}=i
\end{align}
where ${\bm G}_{[{\bm p}_k]}$ denotes the $(N-k)\times(N-k)$ submatrix obtained by deleting the ${\tau}_{0},{\tau}_{1},\ldots{\tau}_{k-1}$-th rows and columns from a matrix whose $ij$-th element is ${G}_{ij[{\bm p}_k]}$. 

We first focus on the $1\times(N-k)$ column vector $\bm g$ corresponding to the original $\tau_k$-th column vector of ${\hat G}_{[{\bm p}_k]}$ such as $\tau_k\neq\tau_0\ldots\tau_{k-1}$ and suppose that ${G}_{\tau_k\tau_k[{\bm p}_k]}$ is located in the $K$-th row and the $1$-th column. Then this is described by the following linear equation 
\begin{align}
(z-{\bm H}_{[{\bm p}_k]}){\bm g}=i{\bm e}
\end{align}
\begin{eqnarray}
{\bm g}=\left(
\begin{array}{c}
{G}_{1\tau_k[{\bm p}_k]}\\
\vdots \\
{G}_{\alpha\tau_k[{\bm p}_k]}\\
\vdots \\
{G}_{N\tau_k[{\bm p}_k]}
\end{array}\right)\hspace{4mm}(\alpha\neq\tau_{0}, \tau_{1},\ldots\tau_{k-1})
\end{eqnarray}
where the $1\times(N-k)$ column vector $\bm e$ is a unit vector whose $K$-th element is 1. 
Therefore the ${G}_{\tau_k\tau_k[{\bm p}_k]}$ in $\bm g$ can be obtained by using Cramer's rule 
\begin{align}
{G}_{\tau_k\tau_k[{\bm p}_k]}=\frac{{\bigr|{\bm F}_{[{\bm p}_{k}]}\bigr|}}{{\bigr|z-{\bm H}_{[{\bm p}_k]}\bigr|}}
\end{align}
where ${\bm F}_{[{\bm p}_{k}]}$ is the $(N-k)\times(N-k)$ matrix obtained by replacing the $K$-th column vector in $z-{\bm H}_{[{\bm p}_{k}]}$ by the column vector $i\bm e$. Using the cofactor expansion along the $K$-th column, the numerator in Eq.(D4) can be rewritten as follows
\begin{align}
{G}_{\tau_k\tau_k[{\bm p}_k]}=i\frac{{\bigr|z-{\bm H}_{[{\bm p}_{k+1}]}\bigr|}}{{\bigr|z-{\bm H}_{[{\bm p}_k]}\bigr|}}
\end{align}
In this way, we can obtain the determinant representation for the effective propagator.
%\begin{acknowledgments}
%\end{acknowledgments}

\bibliography{rereport}

\end{document}